\documentstyle[preprint,aps]{revtex}

\newcommand{\beq}{\begin{equation}}
\newcommand{\eeq}{\end{equation}}
\newcommand{\beqa}{\begin{eqnarray}}
\newcommand{\eeqa}{\end{eqnarray}}
\newcommand{\ket}[1]{| #1 \rangle}
\newcommand{\bra}[1]{\langle #1 |}

\tightenlines

\begin{document}

\draft
\title{An expectation value expansion of Hermitian operators
in a discrete Hilbert space}

\author{Roberth Asplund, Gunnar Bj\" ork\cite{byline},  and Mohamed Bourenanne}

\address {Department of Electronics, Royal Institute of
Technology (KTH), Electrum 229, SE-164 40 Kista, Sweden}

\date{\today}
\maketitle

\begin{abstract}
We discuss a real-valued expansion of any Hermitian
operator defined in a Hilbert space of finite dimension $N$, where
$N$ is a prime number, or an integer power of a prime.
The expansion has a direct interpretation in terms of the operator expectation
values for a set of complementary bases. The expansion can be
said to be the complement of the discrete Wigner function. 

We expect the expansion to be of use in quantum information applications since qubits typically
are represented by a discrete, and finite-dimensional physical system of dimension $N=2^p$, where
$p$ is the number of qubits involved. As a particular example we use the expansion to prove that an intermediate measurement basis (a Breidbart basis) cannot be found if the Hilbert space dimension is 3 or 4.
\end{abstract}


\narrowtext

\section{Introduction}

Quantum physics and information theory are presently melting together in a new interdisciplinary field, quantum information theory. Information science has always centered around discrete, and mainly binary, representations of information. To represent information by discrete states is by no way necessary, but the robustness of discrete (or digital) representations of information against physical imperfections and coupling to the environment has resulted in one of the most rapid and pivotal revolutions of mankind, the digital revolution.

The digital revolution has also affected quantum mechanics, where systems with discrete eigenvalue spectra have recently attracted considerable attention. Examples include few photon states \cite{Bouwmeester}, ion traps \cite{Myatt}, few atom cavity QED systems \cite{Kimble}, superconducting junctions \cite{Nakamura}, nuclear spins \cite{Yannoni}, and Bose-Einstein condensates of atoms \cite{Deng}. With such systems one can implement qubits and qubytes, the quantum analogs to classical binary information.

While quantum mechanics allow feats that seems not be possible in a classical world, such as factorization of numbers in a polynomial time \cite{Shor,Grover}, quantum theory comes with a price. The ubiquitous, and probably natural, representation of quantum mechanics involves complex numbers that makes both computation and visualization of problems more difficult than in a corresponding classical theory that is based on real numbers. In an aptly titled paper, Wootters have addressed the question whether or not complex numbers are strictly necessary in quantum mechanics \cite{Wootters 3}. On this point his paper is inconclusive, we tend to think that the complex numbers are needed. (Recent results by Hardy also indicate that quantum mechanics, describing transitions between states as continuous, naturally leads to a complex number notation \cite{Hardy}.) However, Wootters goes some way towards providing a description of quantum mechanics that avoids, as much as possible, the "phases" of quantum states and expresses the states only in probabilities. Specifically, Wootters proposes a real and positive {\em probability} distribution that is unique for any quantum state defined in a Hilbert space of finite known dimension \cite{Wootters 3}. The probability function can be used as a visualization tool, much like the $P$ \cite{Glauber,Sudarshan}, $Q$, $R$, or Wigner \cite{Wigner} {\em quasi-probability} representations. It can also be used to compute e.g. transition probabilities, but to this end it seems less suitable than the conventional representations. 

In this paper we build on Wootters probability distribution function and show that the function is not only a state representation. It is also directly associated with a state expansion, or more generally, a Hermitian operator expansion due to I. D. Ivanovic \cite{Ivanovic,Note}. This expansion has two appealing features. It has a very symmetric form (it does not rely on a particular basis). Instead, all the complementary (to be defined rigorously below) bases enter on the same footing. In addition, all the expansion coefficients are real and positive, and they equal the measurement probability outcomes of the aforementioned complementary bases. A restriction of the expansion is that it is only defined in Hilbert spaces where the dimension is a prime to an integer power. However, this encompasses the important case $2^p$, $p=1,2, \ldots$, that is, any implementation of qubits.

Wootters has also defined a discrete Wigner function \cite{Wootters} that has many common features, but also has some dissimilarities, to its continuous counterpart. One dissimilarity is that it seems to be more difficult, or at least be less efficient from the measurement point of view, to tomographically reconstruct a system in a Hilbert space of composite dimension, e.g. a spin 5/2 particle that is defined in a $6=2 \cdot 3$ dimensional Hilbert space, than a particle defined in a prime dimensional Hilbert space, e.g. a spin 3 particle, that is defined in a 7 dimensional Hilbert space. We believe the origin of these difficulties can be traced to the underlying discrete mathematics, where the dimension of the linear vector spaces allows, or prohibits, different constructs such as groups, rings and fields. Discrete operator complementarity has an intimate connection to these mathematical constructs \cite{Ivanovic,Wootters 2,Galetti,Vourdas,Ellinas}. Wootters constructed his discrete Wigner function on the same complementary operator set that his probability distribution representation, and our operator expansion, is built on. Recently, we proposed a tomographic reconstruction algorithm for Wootters' Wigner function, based on Ivanovic's expansion \cite{Asplund}. The density matrix, the Wigner function, and Ivanovic's expansion can be seen as three equivalent, closely connected, state representations. In particular, the expansion and the discrete Wigner function can be seen as each other's complement. Going from one to the other involves a rather simple transformation \cite{Asplund}. Tomographic reconstruction of a discrete Wigner function (but defined differently than the one Wootters has proposed) has been discussed by Leonhardt \cite{Leonhardt}. State reconstruction algorithms of states belonging to a discrete and finite Hilbert space have been discussed by several workers \cite{Gale}. Determination of a {\em a priori} unknown state in a $N$ dimensional space will require $N^2-1$ projective measurements if they are non-redundant. If, in addition, they are optimal, all the various projections should either be mutually orthogonal or complementary. An interesting proposal has been put forth by Amiet and Weigert, where they show that reconstruction of a spin state can be accomplished through spin measurements performed by a Stern-Gerlach apparatus oriented along various directions in space \cite{Amiet}. However, we suspect that the projections are not optimal in the sense discussed above.

The paper is organized as follows. The expansion is introduced in Sec. \ref{sec: Expansion}. In Sec. \ref{sec: interm. basis} we apply the expansion to an example. Specifically, we try to find the intermediate measurement basis for eavesdropping applications in quantum secret key distribution. In Sec \ref{sec: conclusions} we summarize our findings. Finally, a proof of the expansion is relegated to Appendix \ref{sec: appendix}. The proof is somewhat tedious, but one point of the proof is worth observing. Most of the proof is void of quantum mechanical ``phases'', but to complete it, we have needed to invoke the phase relationships between states from the different complementary bases. This makes us pessimistic about a formulation of quantum mechanics based entirely on real numbers with no complex ``phases''.

\section{An operator expansion}
\label{sec: Expansion}

Consider an $N$-dimensional Hilbert space ${\cal H}_N$, spanned by
the set of orthonormal basis vectors $\{\ket{\psi_{01}},\ket{\psi_{02}},\ldots,\ket{\psi_{0N}}\}$. The associated density operators will be denoted
$\{\hat{\rho}_{01},\hat{\rho}_{02},\ldots,\hat{\rho}_{0N}\}$.
Since this is a complete basis, there exist an associated
observable, and some physical measurement apparatus corresponding to this
observable can be constructed. Conversely, the eigenvectors of any observable,
defined in this Hilbert space can serve as a complete basis. If
the space dimension $N$ is a prime, or the power of a prime, then
there exist a set of $N$ additional bases, each with
$N$ orthonormal
vectors
$\{\ket{\psi_{k1}},\ket{\psi_{k2}},\ldots,\ket{\psi_{kN}}\}$ where
$k=1,2,\ldots,N$ (the associated density matrices will be denoted
$\hat{\rho}_{kl}$) such that all the $N+1$ bases are mutually
complementary
\cite{Ivanovic,Wootters 2}. Complementary bases (sometimes called mutually unbiased bases) in $\cal{H}_N$ have the property that
\beq
{\rm Tr} \{ \hat{\rho}_{kl} \hat{\rho}_{mn} \} = \left \{
\begin{array}{ll}
\delta_{ln} & \mbox{if $k=m$,} \\
1/N \quad \forall \quad l,n & \mbox{if $k \neq m$.}
\end{array}
\right .
\label{eq: inner product}
\eeq
That is, if the state $\ket{\psi_{kl}}$ is prepared in basis
$k$ but measured in basis $m \neq k$, all $N$ possible
measurement outcomes are equally probable regardless of $l$. This
property is exploited in various quantum key distribution
protocols \cite{DBruss,HBech-NGisin}.

The $N+1$ bases also have an intimate relation to the
discrete Wigner function defined by Wootters
\cite{Wootters}. Alternatively one could claim that it is
Wootter's discrete Wigner function that is related to
complementarity. Indirectly these operators also have connections
to the discrete cyclic permutation groups
\cite{Ivanovic,Galetti,Ellinas}.

Now consider an arbitrary Hermitian operator $\hat{O}$ defined
in ${\cal H}_N$. Ivanovic \cite{Ivanovic} has shown that it is always possible
to expand this operator as
\beq
\hat{O} = \sum_{k=0}^N 
\sum_{l=1}^N {\rm Tr}\{\hat{\rho}_{kl} \hat{O} \} \hat{\rho}_{kl}
 - {\rm Tr}\{ \hat{O} \}\hat{1},
\label{eq:general construction}
\eeq 
where $\hat{1}$ is the identity operator. (Since any operator $\hat{O}$ can be written as the linear combination of the two Hermitian operators $\hat{O}+\hat{O}^\dagger$ and $i [\hat{O}-\hat{O}^\dagger]$ the expansion is in fact valid for any operator although in this case the expansion coefficients ${\rm Tr}\{\hat{\rho}_{kl} \hat{O} \}$ are not real numbers.) We shall derive this result in a different manner in the Appendix. The main difference
between this expansion and the more conventional
expansion 
\beq
\hat{O} = \sum_{q=1}^N 
\sum_{n=1}^N \bra{\psi_{mq}} \hat{O}
\ket{\psi_{mn}} \ket{\psi_{mq}}\bra{\psi_{mn}}
\label{eq:normal construction}
\eeq
is that (\ref{eq:general construction}) only contains real-valued expansion
coefficients ${\rm Tr}\{\hat{\rho}_{kl} \hat{O} \}$. These
coefficients are the respective expectation values of the states
$\ket{\psi_{kl}}$ if measured by the observable $\hat{O}$. In
addition, the expansion (\ref{eq:general construction}) has a ``symmetric'' form, all bases in the
complementary basis sets enter the expansion on equal
footing. Of course there exist an infinite number of
possibilities to choose the complete set of complementary bases
since any unitary transformation will rotate the bases
while preserving their respective mutual properties, and in
particular (\ref{eq: inner product}). However, in (\ref{eq:normal
construction}) only one of the $N+1$ bases is used in the
expansion. For problems with an inherent "symmetry" with respect
of the complementary bases, such as secret key distribution, the
expansion (\ref{eq:general construction}) offers clear
advantages. The ``price'' to be paid by using this expansion is
that the expansion vectors form an overcomplete set and therefore
we must subtract a term proportional to the identity. In addition, the expansion is only defined in Hilbert spaces with a dimension that is a prime to an integer power. This is because there is no known way to construct $N+1$ mutually complementary bases in a $N$ dimensional composite Hilbert space. It is not even known if such a construction is even possible (we suspect it is not).

Consider the standard expansion (\ref{eq:normal construction}) of a general Hermitian operator in ${\cal H}_N$. The operator
is fully described by $N^2$ complex coefficients. However, since
$\hat{\rho}$ is Hermitian, only $N^2$ real numbers are needed to
define the $N^2$ complex coefficients.  The expansion (\ref{eq:general construction}) contains $(N+1)N$ real coefficients. However, since for every basis $k$, the sum of the real coefficients equals the trace of the operator, that is $\sum_{l=1}^N {\rm Tr}\{\hat{\rho}_{kl} \hat{O} \} = {\rm Tr}\{ \hat{O} \}$, we only need to know all $N$ real coefficients in one basis. In the remaining $N$ bases it suffices to know $N-1$ of the coefficients, the last one can be deduced from the trace of the operator. Thus, the expansion (\ref{eq:general construction})
is also fully specified by $N + N(N-1)=N^2$ real numbers. (If the operator happens to be a density operator $\hat{\rho}$ describing a general state, only $N^2-1$ real numbers are needed since ${\rm Tr}\{\hat{\rho} \} = 1 $.) This demonstrates that the expansion (\ref{eq:general construction}) can be considered a maximally
compact representation in the same sense that the expansion (\ref{eq:normal construction}) can be called maximally compact. 

Furthermore, since the coefficient ${\rm
Tr} \{ \hat{\rho}_{kl} \hat{\rho}
\}$ is the projection probability of the state
$\hat{\rho}$ onto the vector $\ket{\psi_{kl}}$,
the expansion shows that if the state $\hat{\rho}$ is unknown, that is, no {\em a priori} information about the expansion coefficients ${\rm Tr} \{ \hat{\rho}_{kl} \hat{\rho} \}$ exist,
the optimal way to estimate them using non-adaptive
von Neumann measurements is to subdivide an ensemble of $M$ identically
prepared systems into $N+1$ sub-ensembles, each containing $M/(N+1)$ members, and on each of the
sub-ensembles measure the statistical distribution of one of the
$N+1$ complementary operators to obtain an estimate of the coefficients. That this is the optimal way to proceed follows from: 1) That a measurement of an ensemble of states in basis $k$ carries no information about the outcome of a measurement on an identically prepared ensemble using basis $m \neq k$ due to (\ref{eq: inner product}). 2) That all expansion coefficients ${\rm Tr} \{ \hat{\rho}_{kl} \hat{\rho} \}$ are needed to fully specify the state. 3) That the minimum number of von Neumann measurements containing $N^2 -1$ mutually orthogonal, or complementary, projections is $N+1$. That such a measurement is an optimal non-adaptive von Neumann \textit{state reconstruction} algorithm was shown earlier by Wootters and Fields \cite{Wootters 2}. They proved that such a measurement strategy minimizes the volume of the Hilbert space in which one can expect to find the state with a given confidence. However, it has also been shown that measurements that give a better fidelity \cite{Peres} or a smaller Bayes cost \cite{Brody} between the estimated state (for a finite measured ensemble) and the prepared state exist. They are either a joint measurement on the whole ensemble of systems \cite{Peres}, or sequential adaptive measurements \cite{Brody,Fisher}.

\section{Construction of an intermediate measurement basis}
\label{sec: interm. basis}

In this section we will demonstrate the utility of the expansion (\ref{eq:general construction}) in problems with an inherent symmetry between a full set of complementary bases. The example we have picked is the construction of an intermediate measurement basis, or a Breidbart basis.

The use of quantum states for fundamentally secure  distribution
of cryptographic secret key distribution was proposed several years
ago. The original proposal, called BB84, defined a two-state, two complementary bases
protocol \cite{Bennett}. An extension made by Bru\ss \cite{DBruss} and Bechmann-Pasquinucci and Gisin \cite{HBech-NGisin} to a six-state (three complementary bases) protocol shows that an eavesdropper's (by convention denoted by Eve) information gain for a given impaired error rate is lower than in the BB84 protocol \cite{DBruss,HBech-NGisin}. Very recently Bechmann-Pasquinucci and Tittel \cite{HBech-WTittel}, Bechmann-Pasquinucci and Peres \cite{HBech-APeres}, and Bourenanne, Karlsson and Bj\" ork \cite{Bourenanne} have considered schemes using four states and two bases, three states and four bases, and $M$ states and $N$ bases, respectively. Below we shall consider an
$N$-state, $N+1$ bases protocol, of which the aforementioned
two-state, three bases protocol is a special case.

In such a quantum secret key distribution system one party (Alice)
randomly chooses a letter $l$ from an $N$-ary
alphabet and a basis $k$ among $N+1$ complementary bases. She subsequently prepares and
transmits the state
$\hat{\rho}_{kl}$ to the other party (Bob). Bob randomly chooses a
measurement basis $m$ among the $N+1$ possibilities, and if the basis coincides with Alice's, he
concludes that the received state is $\hat{\rho}_{kl}$. If he
chooses the wrong base, $m \neq k$, (which will happen with a
probability
$N/(N+1)$), then he will detect the state $\hat{\rho}_{mn}$ with
probability $1/N$ and conclude that this was the received state.
When Alice and Bob publicly compare the used transmission and
detection bases they can discard the events when their respective
bases did not coincide and keep only those when the bases, and
therefore the transmitted and received symbol, coincide.

An intermediate basis \cite{error}, that is a suitable measurement basis for an eavesdropper, utilizes the symmetry properties of the transmission protocol. The intermediate basis for the case of a two state, two bases protocol is illustrated in Fig. \ref{fig: Intermediate}. The idea is to find a measurement operator that lets Eve correctly infer the sent state with high probability. This assures that when she passes along her measurement eigenstate to Bob, the state Alice originally sent, and the state passed along, will have a high fidelity. Hence, Eve gains substantial information about Alice's and Bob's key, while disturbing the states Alice sends to Bob relatively little.

The intermediate operator consists of a set of orthonormal Hermitian operators
$\{\hat{\varrho}_{1},\hat{\varrho}_{2},\ldots,\hat{\varrho}_{N}\}$
such that each operator optimizes the expansion probability of
the states representing one specific letter in the alphabet and
makes the expansion probabilities for the states representing the
other letters equal. Mathematically, this can be stated as
\beq
{\rm Tr} \{ \hat{\varrho}_{l} \hat{\rho}_{k1} \}= \ldots = {\rm
Tr} \{ \hat{\varrho}_{k} \hat{\rho}_{kN} \} = x \quad \forall
\quad l,
\label{eq: Max condition}
\eeq
and
\beq
{\rm Tr}\{\hat{\varrho}_{n}\hat{\rho}_{kl}\}=y \quad \forall
\quad k
\quad {\rm and} \quad
\forall \quad l\neq n ,
\label{eq: Equal condition}
\eeq
where $ 0 < y < x$. 
In addition, we require orthonormality between the intermediate operators
\beq
{\rm Tr} \{ \hat{\varrho}_{n} \} =1 \quad \forall \quad n,
\label{eq:Normalization}
\eeq
\beq
{\rm Tr} \{ \hat{\varrho}_{l} \hat{\varrho}_{n} \}=\delta_{ln}.
\label{eq:Orthogonality}
\eeq

In the example illustrated in Fig. \ref{fig: Intermediate} we can see that ${\rm Tr} \{ \hat{\varrho}_{l} \hat{\rho}_{1l} \}={\rm Tr} \{ \hat{\varrho}_{l} \hat{\rho}_{2l} \}=\cos^2(\pm \pi/8)\approx 0.85$ for $l=1,2$ and ${\rm Tr} \{ \hat{\varrho}_{1} \hat{\rho}_{k2} \}={\rm Tr} \{ \hat{\varrho}_{2} \hat{\rho}_{k1} \} = \cos^2(\pi/2 \pm \pi/8) \approx 0.15$ for $k=1,2$. For a two state, three bases protocol, the construction of an intermediate basis can also be done by a geometrical construction (on the Bloch sphere). In protocols involving only two states in each of $N>3$ complementary bases the construction of the intermediate basis is still relatively straight-forward \cite{Bourenanne}. However, for the protocol we consider here, it has hitherto not been obvious how, or even if it is possible, to make the construction when $N>2$.

To construct these operators we use the expansion (\ref{eq:general
construction}). Using the properties above, and symmetry of the
operator expansion, we get
\beq
\hat{\varrho}_n = \sum_{k=0}^N \left (x \hat{\rho}_{kn} +
\sum_{\stackrel{l=1}{l\neq n}}^N y \hat{\rho}_{kl}
\right ) - \hat{1}.
\label{eq:intermediate construction}
\eeq 
This construction makes $\hat{\varrho}_n$ of a form such that it can
satisfy property (\ref{eq: Max condition}) for an
appropriate choice of
$x$, and it will simultaneously satisfy (\ref{eq: Equal condition}).
The normalization condition (\ref{eq:Normalization}) leads to the
additional equation
\beq
y = {1-x \over N-1}
\label{eq:y}
\eeq
and, finally, inserting (\ref{eq:y}) in the
orthogonality condition (\ref{eq:Orthogonality}) leads to the
relation
\beq
x = {1 \over N} + \sqrt{{1 \over N^2} + {N-3 \over N(N+1)}},
\label{eq:x}
\eeq
where the solution giving $x > y$ is chosen. 

The {\em caveat emptor} of this construction algorithm is that we
are not guaranteed that the intermediate operators derived this way
are von Neumann projection operators, and they may in principle
not even have a representation as a positive operator
measurement. This must be
checked separately.

\subsection {$N$ = 2}

In two-dimensional Hilbert-space, the explicit construction of the
complementary bases is straightforward. Choose any two orthogonal
vectors $\ket{\psi_{01}}$  and $\ket{\psi_{02}}$. The complementary
operator eigenstates can then be chosen to be
\begin{eqnarray}
\ket{\psi_{11}} & = & (\ket{\psi_{01}}+\ket{\psi_{02}})/\sqrt{2}, \\
\ket{\psi_{12}} & = & (\ket{\psi_{01}}-\ket{\psi_{02}})/\sqrt{2},
\end{eqnarray}
and
\begin{eqnarray}
\ket{\psi_{21}} & = & (\ket{\psi_{01}}+i \ket{\psi_{02}})/\sqrt{2},
\\
\ket{\psi_{22}} & = & (\ket{\psi_{01}}-i \ket{\psi_{02}})/\sqrt{2}.
\end{eqnarray}
A physical example of such a triplet of complementary two state
operators are the spin operators in the $x$,
$y$, and $z$-directions of a  spin $1/2$ particle. Plugging these states into
(\ref{eq:intermediate construction}), and using (\ref{eq:y}) and
(\ref{eq:x}), one finds the operators
\begin{eqnarray}
\hat{\varrho}_{1} & = & {1 \over 2 \sqrt{3}} \left [
\begin{array}{cc} 
\sqrt{3}+1 & 1+i \\
1-i & \sqrt{3}-1\end{array}
\right ] , \label{eq:Intermediate 21} \\
\hat{\varrho}_{2} & = & {1 \over 2 \sqrt{3}} \left [
\begin{array}{cc} 
\sqrt{3}-1 & -1-i \\
-1+i & \sqrt{3}+1
\end{array}
\right ] .
\label{eq:Intermediate 22}
\end{eqnarray}
It is easy to check that the two operators have only one
non-zero eigenvalue and hence correspond to a von Neumann measurement by virtue of (\ref{eq:Normalization}) and (\ref{eq:Orthogonality}).
From (\ref{eq:x}) it is clear that if Eve uses this intermediate
basis, she will correctly guess which letter Alice encoded on the
transmitted state with probability $1/2 + 12^{-1/2} \approx 0.789$ irrespective of which of the three bases Alice used for the
encoding. If, instead, she were to chose the simpler intercept and resend strategy, she would only get the right answer with probability $2/3 \approx 0.667$. In\cite{Bourenanne} various eavesdropping strategies for high dimensional secret key distribution protocols are compared.

\subsection {$N$ = 3}

The respective basis vectors in this space can be constructed
using the method in \cite{Ivanovic}. Repeating the construction
algorithm for the intermediate operators for this Hilbert space, one
finds that the operators
are given by
\beq
\hat{\varrho}_{1}  =  {1 \over 6} \left [
\begin{array}{ccc} 
4 & 0 & 0 \\
0 & 1 & 3 \\
0 & 3 & 1
\end{array}
\right ] ,
\label{eq:Intermediate 31} 
\eeq
\beq
\hat{\varrho}_{2}  =  \left [
\begin{array}{ccc} 
 {1 \over 6} & 0 & 0 \\
0 & {2 \over 3} & {-1+ i \sqrt{3} \over 4} \\
0 & {-1- i \sqrt{3} \over 4} & {1 \over 6}
\end{array}
\right ] , 
\label{eq:Intermediate 32} 
\eeq
and
\beq
\hat{\varrho}_{3}  =  \left [
\begin{array}{ccc} 
{1 \over 6}  & 0 & 0 \\
0 & {1 \over 6}  & {-1- i \sqrt{3} \over 4} \\
0 & {-1+ i \sqrt{3} \over 4} & {2 \over 3} 
\end{array}
\right ] . 
\label{eq:Intermediate 33} 
\eeq
Although these operators satisfy all of the criteria (\ref{eq: Max condition}) - (\ref{eq:Orthogonality}) they do not
correspond to any physical measurements because they all have one negative
eigenvalue (the eigenvalues of $\hat{\varrho}_1$ are $-1/3$, $2/3$ and $2/3$, and the eigenvalues of both $\hat{\varrho}_2$ and $\hat{\varrho}_3$ are $(5-3\sqrt{5})/12$, $1/6$, and $(5+3\sqrt{5})/12$). Since
the expansion (\ref{eq:general construction}) is general in the
sense that any Hermitian operator has an unique such expansion, and
(\ref{eq:Intermediate 31}) is the only solution satisfying (\ref{eq:
Max condition}) - (\ref{eq:Orthogonality}), we conclude that an
intermediate measurement basis cannot be found in this space.

\subsection {$N$ = 4}

The result in this case is similar to that of the
three-dimensional Hilbert space. The complementary eigenvectors can
be constructed using the method described in \cite{Wootters 2}. We
find that the first of the intermediate operators is
\beq
\hat{\varrho}_1  =  {1 \over 2 \sqrt{5}} \left [
\begin{array}{cccc} 
 {\sqrt{5}+3 \over 2} & i & 1+i & 0 \\
-i & {\sqrt{5}-1 \over 2} & 1-i & 0 \\
1-i & 1+i & {\sqrt{5}-1 \over 2} & 1 \\
0 & 0 & 1 & {\sqrt{5}-1 \over 2} 
\end{array}
\right ] , 
\label{eq:Intermediate 41} 
\label{eq:z}
\eeq
We do not explicitely give the other three, since all the operators have one negative eigenvalue and therefore cannot correspond to a directly measurable quantity.

\subsection {Conjecture}
From our trials above, we conjecture that for Hilbert-spaces of dimension larger than two, intermediate measurement operators do not exist. It seems that the severe constraints for such an operator, expressed by Eq. (\ref{eq: Max condition})-(\ref{eq:Orthogonality}), do not permit positive Hermitian operators, although Hermitian operators with negative eigenvalues fulfilling the constraints can be found.

\section{Conclusions}
\label{sec: conclusions}

We have discussed an expansion \cite{Ivanovic} of an arbitrary Hermitian operator based on the $N+1$ complementary bases in a Hilbert space of dimension $N$, where $N$ is a prime number, or an integer power of a prime. If one expands a density operator this way, the expansion coefficients constitute a compact probability distribution function with $N(N+1)$ real, positive coefficients, unique to every state, as demonstrated by Wootters \cite{Wootters 3}. We have subsequently used the expansion to try to find an intermediate measurement basis, suitable for eavesdropping in secret key distribution applications. We find that a set of Hermitian operators with the proper characteristics can be found in every Hilbert space of prime, or integer power of a prime, dimension. However, when examining the solutions in Hilbert spaces of dimension 2, 3, and 4, we find that only in the first case does the operators correspond to a physical measurement. From these examples we conjecture that in Hilbert spaces of dimension higher than two, it is not possible to construct an intermediate basis measurement apparatus.

\acknowledgments

The authors would like to thank Mr. Jonas S\"{o}derholm for bringing Ref. \cite{Amiet} to our attention. We also thank Dr. Michael Hall and Professor Bengt Nagel for their comments on an earlier version of this manuscript. The work was supported by grants from the European Commission through the IST FET QIPC QuComm project, the Swedish Research Council for Engineering
Sciences (TFR), and the Swedish Natural Science Research Council (NFR).

\appendix
\section{Proof}
\label{sec: appendix}

To prove that the expansion (\ref{eq:general construction}) holds
for an arbitrary Hermitian operator we expand the operator
according to (\ref{eq:normal construction}) and insert the
expansion in (\ref{eq:general construction}). We denote the
expansion-coefficient $\bra{\psi_{mq}} \hat{O}
\ket{\psi_{mn}} = O_{qn}$. We get
\begin{eqnarray}
\hat{O} & = & \sum_{k=0}^N 
\sum_{l=1}^N {\rm Tr}\{\hat{\rho}_{kl} \sum_{q=1}^N \sum_{n=1}^N
O_{qn}
\ket{\psi_{mq}}\bra{\psi_{mn}} \} \hat{\rho}_{kl}
 - \nonumber \\ 
& & {\rm Tr}\{ \sum_{q=1}^N \sum_{n=1}^N O_{qn}
\ket{\psi_{mq}}\bra{\psi_{mn}} \}\hat{1} .
\label{eq:proof 1}
\end{eqnarray}
The last sum on the right hand side of (\ref{eq:proof 1})
evaluates trivially to 
\beq
{\rm Tr}\{ \sum_{q=1}^N \sum_{n=1}^N O_{qn}
\ket{\psi_{mq}}\bra{\psi_{mn}} \}\hat{1} = \left ( \sum_{q=1}^N
 O_{qq}\right ) \hat{1}
\label{eq:proof 2}
\eeq
To see what the first sum evaluates to we divide it into two.
One sum runs over the density operators of basis $m$, while the
other sum runs over all the remaining $N$ bases' associated
density operators, v.i.z.
\begin{eqnarray}
\sum_{k=0}^N 
\sum_{l=1}^N {\rm Tr}\{\hat{\rho}_{kl} \sum_{q=1}^N \sum_{n=1}^N
O_{qn}
\ket{\psi_{mq}}\bra{\psi_{mn}} \} \hat{\rho}_{kl}
 & = & \nonumber \\ 
\sum_{l=1}^N \sum_{q=1}^N \sum_{n=1}^N
{\rm Tr}\{\hat{\rho}_{ml} O_{qn}
\ket{\psi_{mq}}\bra{\psi_{mn}} \} \hat{\rho}_{ml} +
\nonumber & &\\
\sum_{\stackrel{k=0}{k\neq m}}^N 
\sum_{l=1}^N  \sum_{q=1}^N \sum_{n=1}^N
{\rm Tr}\{\hat{\rho}_{kl} O_{qn}
\ket{\psi_{mq}}\bra{\psi_{mn}} \} \hat{\rho}_{kl} &=& S_1 + S_2 .
\label{eq:proof 3}
\end{eqnarray}
The first term on the right hand side of (\ref{eq:proof 3}) can, using the orthonormality of the basis $m$,
easily be computed.
\beq 
S_1 = \sum_{q=1}^N 
 O_{qq}
 \hat{\rho}_{mq} .
\label{eq:proof 4}
\eeq
The second term we subdivide into
\begin{eqnarray}
S_2 & = & \sum_{\stackrel{k=0}{k\neq m}}^N 
\sum_{l=1}^N  \sum_{q=1}^N O_{qq}
\bra{\psi_{mq}} \hat{\rho}_{kl} 
\ket{\psi_{mq}}  \hat{\rho}_{kl} \nonumber \\
& & + \sum_{\stackrel{k=0}{k\neq m}}^N 
\sum_{l=1}^N  \sum_{q=1}^N \sum_{\stackrel{n=1}{n \neq q}}^N
O_{qn} \bra{\psi_{mn}} \hat{\rho}_{kl} 
\ket{\psi_{mq}}  \hat{\rho}_{kl} \\
&=& S_3 + S_4.
\label{eq:proof 5}
\end{eqnarray}
Using (\ref{eq: inner product}) the term $S_3$ can be reduced to
\beq
S_3 = \sum_{\stackrel{k=0}{k\neq m}}^N 
\sum_{l=1}^N  \sum_{q=1}^N O_{qq} \hat{\rho}_{kl}/N .
\label{eq:one more sum}
\eeq
Noting that $\sum_{l=1}^N \hat{\rho}_{kl}=\hat{1}$ for every $k$,
the sum can be further simplified to
\beq
S_3= N \sum_{q=1}^N O_{qq} \hat{1} /N = \left ( \sum_{q=1}^N
O_{qq} \right ) \hat{1} .
\label{eq:final sum 3}
\eeq
Finally, to simplify $S_4$ we compute the expansion coefficient
$u,v$ of the term containing the term $O_{qn}$ in the basis $m \neq
k$. That is, we delay the summation over $q$ and $n$ until we have found a convenient way to simplify the sums over $k$ and $l$. We find that if $u=q$ and $v=n$ then
\begin{eqnarray}
\bra{\psi_{mu}} \sum_{\stackrel{k=0}{k\neq m}}^N 
\sum_{l=1}^N
\bra{\psi_{mn}} \hat{\rho}_{kl} 
\ket{\psi_{mq}}  \hat{\rho}_{kl} \ket{\psi_{mv}} & =  & \nonumber \\
\bra{\psi_{mq}} \sum_{\stackrel{k=0}{k\neq m}}^N 
\sum_{l=1}^N
\bra{\psi_{mn}} \hat{\rho}_{kl} 
\ket{\psi_{mq}}  \hat{\rho}_{kl} \ket{\psi_{mn}} & =  & \nonumber \\
\sum_{\stackrel{k=0}{k\neq m}}^N 
\sum_{l=1}^N \bra{\psi_{mq}} \psi_{kl} \rangle \langle \psi_{kl}
\ket{\psi_{mq}} \bra{\psi_{mn}} \psi_{kl} \rangle \langle
\psi_{kl} \ket{\psi_{mn}} & = &\nonumber \\
\sum_{\stackrel{k=0}{k\neq m}}^N 
\sum_{l=1}^N 1/N^2 = 1 ,
\label{eq: uqvn}
\end{eqnarray} where (\ref{eq:
inner product}) was used once more. 
If $u=q$ but $v \neq n$ then we get
\begin{eqnarray}
\bra{\psi_{mq}} \sum_{\stackrel{k=0}{k\neq m}}^N 
\sum_{l=1}^N
\bra{\psi_{mn}} \hat{\rho}_{kl} 
\ket{\psi_{mq}}  \hat{\rho}_{kl} \ket{\psi_{mv}} & = & \nonumber \\
\sum_{\stackrel{k=0}{k\neq m}}^N 
\sum_{l=1}^N \bra{\psi_{mq}} \psi_{kl} \rangle \langle \psi_{kl}
\ket{\psi_{mq}} \bra{\psi_{mn}} \psi_{kl} \rangle \langle
\psi_{kl} \ket{\psi_{mv}} & = & \nonumber \\
\sum_{\stackrel{k=0}{k\neq m}}^N 
\sum_{l=1}^N {\bra{\psi_{mn}} \psi_{kl} \rangle \langle
\psi_{kl} \ket{\psi_{mv}} \over N} = \sum_{\stackrel{k=0}{k\neq
m}}^N 
{\bra{\psi_{mn}} \hat{1} \ket{\psi_{mv}} \over N} & = & 0,
\label{eq: unqvnn}
\end{eqnarray}
The same result is obtained when $u \neq q$ but $v = n$. Note that so far in our proof we have not needed to invoke any ``phase relationship'' between the complementary states. However, to complete our proof we now need to use the inner product between the complementary basis states including the  ``phases'' of the inner products. Below we shall show that when $u \neq q$ and $v \neq n$ the sum evaluates to
\begin{eqnarray}
\sum_{\stackrel{k=0}{k\neq m}}^N 
\sum_{l=1}^N \bra{\psi_{mu}} \psi_{kl} \rangle \langle \psi_{kl}
\ket{\psi_{mq}} \bra{\psi_{mn}} \psi_{kl} \rangle \langle
\psi_{kl} \ket{\psi_{mv}} = 0 .
\label{eq: phases}
\end{eqnarray}
Since Eq. (\ref{eq: uqvn})-(\ref{eq: phases}) are true
for every $q$ and $n$ in the sum $S_4$, we can
use the following simplification before summing $S_4$ over the indices $\sum_{\stackrel{k=0}{k\neq m}}^N 
\sum_{l=1}^N O_{qn}
\bra{\psi_{mn}} \hat{\rho}_{kl} 
\ket{\psi_{mq}}  \hat{\rho}_{kl} = O_{qn}
\ket{\psi_{mq}}\bra{\psi_{mn}}$. The sum $S_4$ now can be expressed
\beq
S_4  = \sum_{q=1}^N
\sum_{\stackrel{n=1}{n \neq q}}^N O_{qn}
\ket{\psi_{mq}}\bra{\psi_{mn}} .
\label{eq: sum 4}
\eeq 
Finally, summing (\ref{eq:proof 4}), (\ref{eq:final sum 3}),
(\ref{eq: sum 4}) and subtracting (\ref{eq:proof 2}), the
standard expansion (\ref{eq:normal construction}) is obtained. The
proof is hence completed.

\subsection{Proof of (\ref{eq: phases}) when $N$ is a power of an odd prime}

Here we prove that (\ref{eq: phases}) is correct provided that the Hilbert space dimension $N=N_i^p$ is a power of an odd prime, $q \neq u$, and $v \neq n$.

We follow the nomenclature in \cite{Wootters 2} to express the phases occurring in the scalar products in (\ref{eq: phases}). To this end we assign a $p$-component column vector with entries corresponding to the index value expressed in base $N_i$ to each of the basis vector indices $u$, $l$, $q$, $n$, and $v$. E.g if $N=81=3^4$ the value $u=71=2 \cdot 3^3 + 1 \cdot 3^2 + 2 \cdot 3 + 2$ has the associated vector ${\bf u}^T= (2,1,2,2)$, where the superscript $T$ indicates the transpose. A $p$-component column vector ${\bf k}$ is also associated to each basis, excluding the basis Wootters and Fields calls the ``standard basis'', which we will take to be the base $m$. Since the first sum in Eq. (\ref{eq: phases}) sum runs over all bases other than our chosen standard basis, the numbering of the remaining $N$ bases is irrelevant. The vectors ${\bf k}^T$ hence run from $(0,0,\ldots,0)$ to $(N_i -1,N_i -1,\ldots,N_i -1)$. Finally we use the tensor ${\bf \alpha}$ that is a $p$-component vector of $p$ by $p$ matrices defining the field algebra. Every matrix ${\bf \alpha}_j$, where $j=0,1,\ldots,p$, is a symmetric matrix. Now assume that $N_i \neq 2$ and use the complementary bases defined by Wootters. The scalar products in (\ref{eq: phases}) becomes \cite{Wootters 2}
\begin{eqnarray}
\frac{1}{N^2}
\sum_{k=1}^N
\sum_{l=1}^N 
e^{(2\pi i /N_i)[{\bf u}^T({\bf k \cdot \alpha}){\bf u}+{\bf l}^T {\bf u}]}
e^{-(2\pi i /N_i)[{\bf q}^T({\bf k \cdot \alpha}){\bf q}+{\bf l}^T {\bf q}]} \cdot \nonumber \\
e^{(2\pi i /N_i)[{\bf n}^T({\bf k \cdot \alpha}){\bf n}+{\bf l}^T {\bf n}]}
e^{-(2\pi i /N_i)[{\bf v}^T({\bf k \cdot \alpha}){\bf v}+{\bf l}^T {\bf v}]}=\nonumber \\
\frac{1}{N^2}
\sum_{k=1}^N
e^{(2\pi i /N_i)[{\bf u}^T({\bf k \cdot \alpha}){\bf u}-{\bf q}^T({\bf k \cdot \alpha}){\bf q}} \cdot \nonumber \\
e^{(2\pi i /N_i){\bf n}^T({\bf k \cdot \alpha}){\bf n}-{\bf v}^T({\bf k \cdot \alpha}){\bf v}]}
\sum_{l=1}^N  e^{(2\pi i /N_i)[{\bf l}^T ({\bf u}-{\bf q}+{\bf n}-{\bf v})]} .
\label{eq: product sum}
\end{eqnarray}
The last sum in this expression will be $N$ if and only if $u-q+n-v=0$. Otherwise 
the result will be zero. However, we are interested only in the case when $n \neq q$. Let us put $n=q+s$, where $s \neq 0$. The equation above then gives $v=u+s$. Inserting these relations, and defining a $p$-component column vector ${\bf s}$ in analogy with the other index vectors, leads to

\begin{eqnarray}
\frac{1}{N}
\sum_{k=1}^N
e^{(2\pi i /N_i)[{\bf u}^T({\bf k \cdot \alpha}){\bf u}-{\bf q}^T({\bf k \cdot \alpha}){\bf q}} \cdot \nonumber \\
e^{(2\pi i /N_i)({\bf q}^T + {\bf s}^T)({\bf k \cdot \alpha})({\bf q}+{\bf s})-({\bf u}^T + {\bf s}^T)({\bf k \cdot \alpha})({\bf u}+{\bf s})]} = \nonumber \\
\frac{1}{N} \sum_{k=1}^N
e^{(2\pi i /N_i)[({\bf q}^T-{\bf u}^T)({\bf k \cdot \alpha}){\bf s}+{\bf s}^T ({\bf k \cdot \alpha})({\bf q}-{\bf u})]}
\label{eq: uqs sum}
\end{eqnarray}
At this point we recall the meaning of the tensor product in the exponent of (\ref{eq: uqs sum}). We have defined a $N$ element field. We can number the field elements with an index running from $1$ to $N$, and this number can be expressed as a p-component vector if $N-i$ is used as the basis, as exemplified with $u$ above. The product between any two elements of the field is also an element of the field. Suppose we multiply the field elements number $q-u$ and $s$. We assume that the result is field element number $r$. If $r$ is expressed as a vector ${\bf r}$, then the $(p-j)$th coefficient of ${\bf r}$ will be $({\bf q}^T-{\bf u}^T){\bf \alpha}_j{\bf s}$. Multiplication between elements of a field is commutative. Therefore, for any given ${\bf k}$, the following equality holds
\beq
({\bf q}^T-{\bf u}^T)({\bf k \cdot \alpha}){\bf s}={\bf s}^T ({\bf k \cdot \alpha})({\bf q}-{\bf u}) .
\label{eq: commutation}
\eeq
This implies, as stated above, that every matrix ${\bf \alpha}_j$ is symmetric. The relation (\ref{eq: commutation}) enables us to write the sum (\ref{eq: uqs sum}) as
\begin{equation}
\frac{1}{N}
\sum_{k=1}^N
e^{(4\pi i /N_i)[({\bf q}^T-{\bf u}^T)({\bf k \cdot \alpha}){\bf s})]} .
\label{eq: uqs sum 2}
\end{equation}
Since we know that $q \neq u$ and $s \neq 0$, the product of the element number $q-u$ and $s$ is a non-zero element of the field. Let us once more assume that the product is the $r$th element. Forming the tensor product $({\bf q}^T-{\bf u}^T)({\bf k \cdot \alpha}){\bf s}$, and stepping the index vector ${\bf k}$ successively through all the numbers from $1$ to $N=N_i^p$, will multiply each component of the vector ${\bf r}$ an equal number of times with the integers $0, 1, \ldots, N_i-1$, and then add the results, in such a way that the shortest repetitive cycle of the whole process is exactly $N$. The digit sum [modulus $N_i$, but this is already naturally incorporated in (\ref{eq: uqs sum 2}) through the pre-factor $(4\pi i /N_i)$ in the exponent] will equal each of the numbers $0, 1, \ldots, N_i$ exactly $N/N_i=N_i^{p-1}$ times, and therefore the sum (\ref{eq: uqs sum 2}) will equal zero. 

To fully appreciate what was just said we will exemplify in Table \ref{table: 1} the calculations by assuming that $N=9$ giving $N_i=3$ and $p=2$. We will demonstrate the statement above assuming that the product of the elements number $q-u$ and $s$ is $r=1 \rightarrow {\bf r}_1=(0,1)$, $r=4 \rightarrow {\bf r}_4=(1,1)$, and $r=5 \rightarrow {\bf r}_5=(1,2)$, respectively. The assertion is of course true for any $r=1,2,\ldots,9$. In the three ``Sum'' columns we see that each of the integers 0, 1, and 2 appears exactly $N/N_1=9/3=3$ times. Hence, the sum of the exponentials (\ref{eq: uqs sum 2}) will equal zero. It is not hard to see why this will be true for any combination of non-zero elements $q-u$ and $s$.

\subsection{Proof of (\ref{eq: phases}) when $N$ is a power of two}

If $N_i=2$, the proof goes along the same lines and one arrives at the sum
\begin{equation}
\frac{1}{N}
\sum_{k=1}^N
e^{(\pi i)[({\bf q}^T-{\bf u}^T)({\bf k \cdot \alpha}){\bf s})]} ,
\label{eq: uqs sum 3}
\end{equation}
instead of (\ref{eq: uqs sum 2}), where, again, $q-u \neq 0$ and $s \neq 0$. By the construction of the field algebra, the exponent will take on the values 0 and 1, respectively, exactly $2^{p-1}=$ times as $k$ goes from 1 to $N=2^p$. Therefore the sum equals zero in this case, too.

\begin{figure}
\caption{In the BB84 secret key distribution protocol symbol 1 is randomly coded on state $\ket{\psi_{11}}$ or $\ket{\psi_{21}}$ and symbol 2 is randomly coded on state $\ket{\psi_{12}}$ or $\ket{\psi_{22}}$. The intermediate operator eigenvectors are denoted $\ket{\beta_1}$ and $\ket{\beta_2}$.}
\label{fig: Intermediate}
\end{figure}

\begin{table}
\caption{A table demonstrating the construction of the exponents in Eq. (\ref{eq: uqs sum 2}) for the case $N=9$. The element $r$, indicating the result of the product between the field elements number $q-u$ and $s$, can be any non-zero element of the field. We have, arbitrarily, made the calculations giving the results if $r=1, 4$, and $5$, respectively. We have made the calculations modulus 3, but this is not necessary since the Eq. (\ref{eq: uqs sum 2}) naturally incorporates this by the exponent pre-factor $4 \pi i/3$.}
\begin{tabular}{ccccccccccc}
k&${\bf k}^T$&${\bf r}_1^T$&${\bf k}^T {\bf r}_1$&Sum&${\bf r}_4^T$&${\bf k}^T {\bf r}_4$&Sum&${\bf r}_5^T$&${\bf k}^T {\bf r}_5$&Sum\\
\tableline
1&(0,1)&(0,1)&0+1&1&(1,1)&0+1&1&(1,2)&0+2&2\\
2&(0,2)&(0,1)&0+2&2&(1,1)&0+2&2&(1,2)&0+1&1\\
3&(1,0)&(0,1)&0+0&0&(1,1)&1+0&1&(1,2)&1+0&1\\
4&(1,1)&(0,1)&0+1&1&(1,1)&1+1&2&(1,2)&1+2&0\\
5&(1,2)&(0,1)&0+2&2&(1,1)&1+2&0&(1,2)&1+1&2\\
6&(2,0)&(0,1)&0+0&0&(1,1)&2+0&2&(1,2)&2+0&2\\
7&(2,1)&(0,1)&0+1&1&(1,1)&2+1&0&(1,2)&2+2&1\\
8&(2,2)&(0,1)&0+2&2&(1,1)&2+2&1&(1,2)&2+1&0\\
9&(0,0)&(0,1)&0+0&0&(1,1)&0+0&0&(1,2)&0+0&0
\end{tabular}
\label{table: 1}
\end{table}

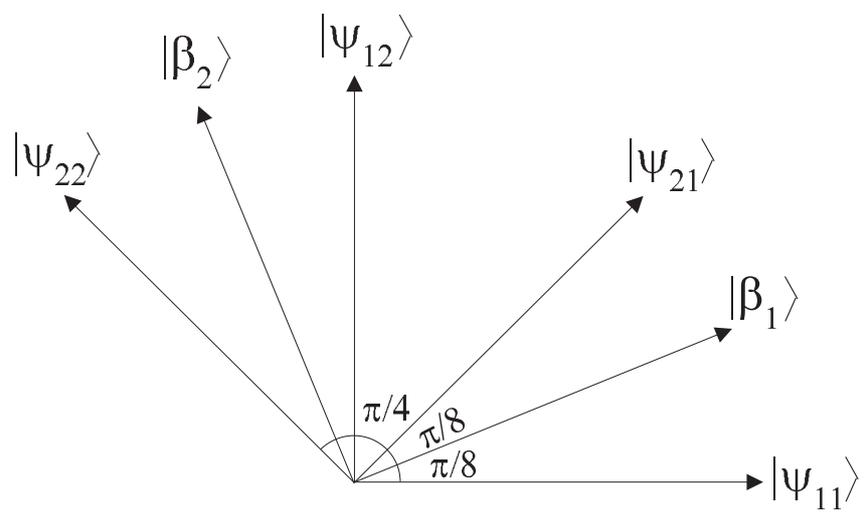
\end{document}